\documentclass[iop]{emulateapj}
\usepackage{auto-pst-pdf}
\usepackage{subfigure}
\usepackage{amsmath, amsthm, amssymb}
\shorttitle{RUN DMC}
\shortauthors{B. Nelson et al.}

\begin{document}

\title{RUN DMC: An efficient, parallel code for analyzing Radial Velocity Observations using N-body Integrations and Differential Evolution Markov chain Monte Carlo}

\author{Benjamin E. Nelson\altaffilmark{1,2,3}, 
Eric B. Ford\altaffilmark{1,2,3}}
\author{Matthew J. Payne\altaffilmark{3,4}}

\email{benelson@psu.edu}

\altaffiltext{1}{Center for Exoplanets and Habitable Worlds, The Pennsylvania State University, 525 Davey Laboratory, University Park, PA, 16802, USA}
\altaffiltext{2}{Department of Astronomy and Astrophysics, The Pennsylvania State University, 525 Davey Laboratory, University Park, PA 16802, USA}
\altaffiltext{3}{Department of Astronomy, University of Florida, 211 Bryant Space Science Center, Gainesville, FL 32611, USA}
\altaffiltext{4}{Harvard-Smithsonian Center for Astrophysics, 60 Garden Street, Cambridge, MA 02138, USA}


\begin{abstract}
In the 20+ years of Doppler observations of stars, scientists have uncovered a diverse population of extrasolar multi-planet systems. 
A common technique for characterizing the orbital elements of these planets is Markov chain Monte Carlo (MCMC), using a Keplerian model with random walk proposals and paired with the Metropolis-Hastings algorithm. 
For approximately a couple of dozen planetary systems with Doppler observations, there are strong planet-planet interactions due to the system being in or near a mean-motion resonance (MMR). 
An N-body model is often required to accurately describe these systems.
Further computational difficulties arise from exploring a high-dimensional parameter space ($\sim$7 x number of planets) that can have complex parameter correlations, particularly for systems near a MMR. 
To surmount these challenges, we introduce a differential evolution MCMC (DEMCMC) applied to radial velocity data while incorporating self-consistent N-body integrations. 
Our Radial velocity Using N-body DEMCMC (RUN DMC) algorithm improves upon the random walk proposal distribution of the traditional MCMC by using an ensemble of Markov chains to adaptively improve the proposal distribution. 
RUN DMC can sample more efficiently from high-dimensional parameter spaces that have strong correlations between model parameters. 
We describe the methodology behind the algorithm, along with results of tests for accuracy and performance. 
We find that most algorithm parameters have a modest effect on the rate of convergence.
However, the size of the ensemble can have a strong effect on performance.
We show that the optimal choice depends on the number of planets in a system, as well as the computer architecture used and the resulting extent of parallelization. 
While the exact choices of optimal algorithm parameters will inevitably vary due to the details of individual planetary systems (e.g., number of planets, number of observations, orbital periods and signal-to-noise of each planet), we offer recommendations for choosing the DEMCMC algorithm's algorithmic parameters that result in excellent performance for a wide variety of planetary systems.

\end{abstract}


\section{Introduction}
Over the past couple of decades, a wide variety of planets have been detected orbiting other stars in the Galactic neighborhood. 
The first wave of exoplanets discovered from radial velocity (RV) surveys were limited to those massive enough and close enough to their host stars to be detected given our technological and temporal capabilities. 
The increasing timespan of observations, as well as improvements in Doppler measurement precision, have enabled RV campaigns to uncover a population of multiple planet systems. 
Today, Doppler surveys continue to become sensitive to lower planet masses and planets in wider orbits. 
Dynamical investigations to understand the formation and evolution of these systems and to test planet formation theories rely on accurately characterizing the range of orbital elements and masses of planets consistent with the observations.

Inferring model parameters (e.g., orbital period, mass, eccentricity, etc.) from RV observations can be quite challenging. 
To fully describe the orbital properties of a planetary system requires 7 dimensions per planet (3 position components, 3 velocity components, and mass). 
As we only observe the line-of-sight component of a star's motion, RV observations can result in near degeneracies amongst model parameters (e.g. mass vs. on-sky inclination, $i$). 
It is difficult to pin down quantities such as $i$, or the ascending node, $\Omega$, without assistance from other observing techniques or strong mutual planetary interactions. 
Even if we consider only coplanar systems, the parameter space contains 5 dimensions per planet, plus one for inclination, plus one for the offset in the velocity zero-point for each observatory/instrument, and at least one stellar/systematic ``jitter'' parameter. 
Therefore, a coplanar system of $n_p$ planets requires exploring a parameter space with a number of dimensions $n_{dim}\ge3 + 5n_p$.

In this paper, we focus on such coplanar systems. For $n_p\ge2$, the larger $n_{dim}$ is much more likely to lead to correlations or degeneracies between model parameters and create a complex $\chi^2$ surface, both of which make it much more challenging to efficiently explore parameter space.

While some studies report best-fit model parameters and uncertainties, the potential for correlated parameters makes it particularly important to characterize the distribution of model parameters consistent with the observations for multi-planet systems. 
In a Bayesian framework, this corresponds to sampling from the posterior probability distribution (\S\ref{secMcmc}). 
Previous studies have developed efficient MCMC-based methods for characterizing systems with one to a few planets, assuming each follow a Keplerian orbit, i.e. neglecting planet-planet interactions \citep{Ford06, Gregory11, Hou12}. 
However, these methods become computationally impractical for systems with strongly interacting planets.

In this paper, we present RUN DMC, a code for analyzing radial velocity observations using N-body integrations and differential evolution Markov chain Monte Carlo.  
RUN DMC combines the differential evolution Markov chain Monte Carlo (DEMCMC; \citet{terBraak06}) algorithm for posterior sampling with N-body integrations to accurately model radial velocity observations of strongly interacting planetary systems.  
DEMCMC is a variation on the standard MCMC algorithm that specializes in navigating high-dimensional parameter spaces including those with correlations among multiple model parameters that would significantly reduce the sampling efficiency of traditional MCMC algorithms.  
To perform the required N-body integrations and computations efficiently, RUN DMC is parallelized for both multi-core workstations and graphics processing units (GPUs).  
We apply the new algorithm to real and simulated RV observations.   
We test the robustness of our algorithm for analyzing observations of planetary systems with a varying number of planets in two ways: 
(1.) efficiency of finding the mode of the (presumed) global minimum of parameter space from an intentionally poor set of initial guesses, and 
(2.) estimating the autocorrelation length of the model parameters once RUN DMC has found this global mode.  

We report the results of these tests applied to various synthetic multi-planet systems.  
In \S\ref{secModel}, we discuss our model of observations, orbital model, and the assumed priors.  
\S\ref{secMcmc} provides and overview for a traditional MCMC algorithm applied to RV observations.  
\S\ref{secDemcmc} describes the DEMCMC algorithm and the specific input parameters of RUN DMC.  
\S\ref{secPerform} explains our testing procedure and the results when tested for a wide variety of synthetic multi-planet systems.  
Finally, in \S\ref{secDiscuss}, we provide general recommendations for using a DEMCMC algorithm on RV data and discuss the implications for analyzing planetary systems with two to five (or more) planets.
While one should always perform convergence tests of Markov chain algorithms before making inferences for specific planetary systems, experience has shown that these recommendations for algorithmic parameters provide an excellent starting point and typically result in sufficiently rapid convergence that the first simulations can be used for inference.
Table \ref{tbl-1} references all the common notation we plan to use in this paper.

\begin{deluxetable}{cc}
\tablecaption{\small Notation and descriptions of selected variables used in this paper.\label{tbl-1} }
\tablewidth{0pt}
\tablehead{ \colhead{Parameter} & \colhead{Description} }
\startdata
$n_p$ & $\textup{number of planets}$ \\
$n_{dim}$ & $\textup{number of dimensions}$ \\
$n_{chains}$ & $\textup{number of Markov chains}$ \\
$n_{gen}$ & $\textup{number of generations}$ \\
$n_{eval}$ & $\textup{number of model evaluations}$ \\
$\vec{x}_n$ & $\textup{\textit{n}th state of a chain}$ \\       
$\vec{x}_n^{\prime}$ & $\textup{proposed \textit{n+1} state of a chain}$ \\     
$\gamma$ & $\textup{vector scale length}$ \\
$\sigma_\gamma$ & $\textup{randomness parameter for } \gamma$ \\                
$\alpha$ & $\textup{scale parameter for scattering perturbation}$ \\
$\beta$ & $\textup{scale parameter for shifting perturbation}$ \\
\textup{AC}$^{-1}(0)$ & $\textup{minimum generation lag when autocorrelation $\approx$ 0}$ \\
\enddata
\end{deluxetable}

\section{Physical \& Statistical Model}
\label{secModel}

\subsection{Physical Model}
\label{secModelPhys}
For the majority of the discovered exoplanet systems, mutual planetary interactions are weak and the baseline of the available RV observations is not long enough to detect dynamical interactions. 
Thus, the perturbation of each planet on its host star can be well approximated using a Keplerian model. In the case of most multi-planet systems, we can approximate the star's motion at any given time by summing over the RV signals due to gravitational perturbations by each planet. 
The input set of model parameters, $\vec{\theta}$, includes orbital period ($P$), RV semi-amplitude ($K$), eccentricity ($e$), longitude of periastron ($\omega$)\footnote{Since all the systems considered in this paper are coplanar, we set the longitude of ascending node, $\Omega$, to 0 without loss of generality.}, and initial mean anomaly ($M$) for each planet.

Assuming no self-interactions amongst the planets, the net velocity perturbation, $v_{\star, \vec{\theta}}(t,j)$, of a star at any given time, $t$, measured at observatory $j$ is given by
\begin{equation}
v_{\star,\vec{\theta}}(t,j)=\sum_{i} K_i\left\{\cos\left[\omega_i+f_i(t)\right]+e_i\cos\omega_i\right\} + C_j
\end{equation}
where the subscript $i$ refers to the $i$th planet and $f(t)$ is the true anomaly. 
The relation between $M$ and $f$ can solved using Kepler's equation and equation 2.46 of \citep{MurrayandDermott}.

In addition to the orbital parameters, the RV model includes velocity offsets, $C_j$, where $j$ denotes which instrument/template was used. 
These arise since high precision RV observations are differential measurements and each instrument uses a different spectral template in the data reduction process.

This paper addresses the issue of multi-planet systems with noticeably strong dynamical interactions on the observing timescale. 
RV surveys have found a couple of dozen systems with multiple planets near a mean-motion resonance or in tightly compact configurations that necessitate a model that accounts for planet-planet interactions. 
Systems such as  GJ 876 \citep{Rivera10}, HD 200964 \citep{Johnson11}, and dozens of recently confirmed Kepler systems \citep{Holman10, Lissauer11, Cochran11, Fabrycky12, Ford12, Steffen12a, Steffen12b} require self-consistent N-body integrations to accurately model the observations. 
The induced gravitational acceleration on the $i$th body from all other bodies is simply,
\begin{equation}
\frac{d^2\vec{r_i}}{dt^2}=-\sum^N_{j=1}\frac{Gm_j(\vec{r_i}-\vec{r_j})}{|\vec{r_i}-\vec{r_j}|^3},
\end{equation}
where $G$ is the gravitational constant, $m_j$ is the mass of the $j$th body, and $\vec{r}$ is the position vector to each body relative to some arbitrary origin. 
The N-body integrations are performed using units of solar mass ($M_\odot$), AU, and $G$ = 1, so one year equals $2\pi$ time units. 
RUN DMC performs these integrations using a time-symmetric 4th order Hermite integrator \citep{Kokubo98b} for a good balance between speed and accuracy. 
We set the integration timestep to no more than 0.5\% of the inner-most orbital period for our simulations, a value recommended through the work of \citet{Kokubo98b}.

\subsection{Model of Observations}
\label{secModelObs}
Spectroscopic observations measure the line-of-sight velocity of a star as it is gravitationally perturbed by orbiting companions, causing periodic variations when observed over time. 
Each spectrum can be reduced into a single velocity measurement, $v_{\star, obs}$, that has a quantifiable measurement uncertainty, $\sigma_{\star, obs}$. 
More specifically, most echelle based RV surveys use the shift of the line profiles of thousands of spectral lines to make a velocity estimation. 
Therefore, the uncertainties in these measurements are nearly Gaussian \citep{Butler96}. 
We also include a ``jitter'' term, $\sigma_{jit}$, that accounts for any unmodelled systematics or astrophysical noise sources (e.g. starspots, pulsations, p-modes). 
In the case of exoplanet surveys, $\sigma_{jit}$ is at least partially due to (and sometimes dominated by) variations in stellar activity, stellar variability, or undetected planets.

\subsection{Likelihood}
\label{secL}
Assuming the uncertainties from individual observations closely follow a Gaussian distribution and are uncorrelated, we can evaluate the goodness of fit to our set of observations, $\vec{d} = \left\{v_{\star, obs}, \sigma_{\star, obs} \right\}$, utilizing the $\chi^2$ statistic. 
\begin{equation}
\chi^2=\sum_k \frac{[v_{\star, obs}(t_k,j_k)-v_{\star, \vec{\theta}}(t_k,j_k)]^2}{(\sigma_{\star, obs}(t_k,j_k)^2+\sigma_{jit}^2)}
\end{equation}
Correlations may arise amongst multiple measurements taken in one night, but in this study, we consider synthetic RV measurements based on real RV time series and are generated independently, so our aforementioned assumption is valid. For a general multi-planet system, $v_{\star, \vec{\theta}}$ is solved using a self-consistent N-body model. 
With an N-body model, each evaluation of $\chi^2$ becomes much more computationally demanding than for a single planet system which can be described by a Keplerian orbit or a multiple planet system that can be well approximated by a linear superposition of Keplerian orbits.

If we assume that each measurement was made independently, we can construct the appropriate {\em likelihood function} for obtaining a set of observations ($\vec{d}$) given the parameter values for a specified model ($\vec{\theta}$).
\begin{multline}
L(\vec{\theta})=p(\vec{d}|\vec{\theta}) = \left[ \prod_k \frac{1}{\sqrt{2\pi(\sigma_{\star, obs}(t_k,j_k)^2+\sigma_{jit}^2)}}\right] \\ \times \exp\left(-\chi^2/2\right).
\end{multline}
Note that we consider the jitter ($\sigma_{jit}$) to be a model parameter to be estimated from the data.  
For computational convenience, we employ an effective chi-squared, so $L(\vec{\theta}) \propto \exp(-\chi^2_{\mathrm eff}/2)$, where $\chi_{eff}^2$ is calculated by
\begin{equation}
\chi_{eff}^2=\chi^2 + \sum_k \ln\left[\frac{\sigma_{\star, obs}(t_k,j_k)^2+\sigma_{jit}^2}{\sigma_{\star, obs}(t_k,j_k)^2}\right].
\end{equation}
Increasing $\sigma_{jit}$ reduces $\chi^2$, but simultaneously increases the right-most term that can be viewed as the natural Bayesian penalty term for large jitter.  

\subsection{Priors}
\label{secPrior}
We adopt broad and separable priors for all of our model parameters.  
\begin{multline}
p(\vec{\theta}) = p(C) p(\sigma_{jit}) \\ \times \prod_i p(P_i) p(K_i) p(e_i) p(\omega_i) p(i_i) p(\Omega_i) p(M_i)
\end{multline}
Note the subscript $i$ denotes which planet, while $p(i_i)$ is the prior probability for the inclination of the $i$th planet.
Table \ref{tbl-2} lists our choice of priors for each parameter which closely follow the SAMSI reference priors \citep{Ford07}, but with the addition of uniform priors for the longitude of ascending node ($\Omega$) and an isotropic prior for the orbital inclination relative to the sky plane ($i$).  

\begin{deluxetable*}{cccc}
\tablecaption{\small The set of model parameters commonly used in the analysis of RV datasets and the assumed prior probability distributions. \label{tbl-2}}
\tablewidth{0pt}
\tablehead{ \colhead{Parameter} & \colhead{Prior} & \colhead{Bounds} }
\startdata
$\textup{Period}, P, \left(P_0 = 1 \textup{day}\right)$ & $p(P) \propto \left(1+P/P_0\right)^{-1}$ & $[0, \infty]$ \\ 
$\textup{RV Amplitude}, K, \left(K_0 = 1 \textup{m/s}\right)$ & $p(K) \propto \left(1+K/K_0\right)^{-1}$ & $[0, \infty]$ \\
$\textup{Mean Anomaly}, M$ & $p(M)=\frac{1}{2\pi}$ & $[0, 2\pi]$   \\
$\textup{Eccentricity}, e$ & $p(e) = 1$ & $[0, 1]$ \\
$\textup{Longitude of Pericenter}, \omega$ & $p(\omega)=\frac{1}{2\pi}$ & $[0, 2\pi]$   \\
$\textup{Longitude of Ascending Node}, \Omega$ & $p(\Omega)=\frac{1}{2\pi}$ & $[0, 2\pi]$   \\
$\textup{Inclination}, i$ & $p(i) \propto \sin{i}$ & $[0, \pi]$    \\
$\textup{Velocity Offset}, C$ & $p(C)=\frac{1}{2C_{\max}}$ & $[-\infty, \infty]$   \\
$\textup{Jitter}, \sigma_{jit}, \left(\sigma_{jit_0}=1 \textup{m/s}\right)$ & $p(\sigma_{jit}) \propto \left(1+\sigma_{jit}/\sigma_{jit_0}\right)^{-1}$ & $[0, \infty]$ \\
\enddata
\end{deluxetable*}

As the exoplanet catalog grows, astronomers hope to learn about the intrinsic distribution of masses, orbital parameters, and the correlations amongst them.  
With almost 900 confirmed planets and more than 3000 planet candidates discovered by the Kepler Space Telescope, astronomers are finding that planetary systems are extremely diverse in orbital architecture.  
In principle, knowledge of the distribution of planet masses and orbits from large surveys could inform priors used to analyze individual systems.  
However, astronomers are still trying to understand the intrinsic distribution of planet masses and orbital parameters.  
We choose to use broad priors, so that the analyses of individual systems can be readily compared or combined for future hierarchical Bayesian analyses.  
Posterior samples generated using RUN DMC and these broad priors can be transformed into posterior samples using a more informative set of priors via either reweighting or rejection sampling, e.g., \cite{Hogg12}.

\subsubsection{Modified Jeffreys Priors}
For most model parameters, we use simple, non-informative priors.  
The priors for the three parameters ($P$, $K$, $\sigma_{jit}$) are modified Jeffreys priors that deserve further explanation.  
A Jeffreys prior ($p(x)\propto x^{-1}; x>0$) is a non-informative prior, with a feature that it is invariant under re-scaling of the model parameter $x$.
This often works well for scale parameters, where the width of a distribution scales with its associated parameter.
However, the traditional Jeffreys prior would result in a physically unrealistic divergence for $P$, $K$, and $\sigma_{jit}$ at small values.  
The prior-induced singularity near zero would result in a highly multi-modal posterior with peaks at physically unrealistic values (e.g., planets inside the star) or corresponding to values for which there is no empirical evidence (e.g., planets with infinitesimal mass).  
Therefore, we impose a modified Jeffreys prior for $P$, $K$ and $\sigma_{jit}$.  
For example, the modified Jeffreys prior for $\sigma_{jit}$ is 
\begin{equation}
p(\sigma_{jit})\propto \left\{\begin{matrix}
 \frac{1}{\left(1+\sigma_{jit}/\sigma_{jit_0}\right)} & \textup { for } \sigma_{jit}>0
 \\ 
0  & \textup { for } \sigma_{jit}\le0
\end{matrix}\right.
\end{equation}
where we set $\sigma_{jit_0}$ to 1 m/s.  
Similarly, we use $K_0=1$ m/s and $P_0=1$ day.
For Bayesian model selection, one would need to include a factor of $1/\ln\left(\frac{1+\sigma_{jit, max}/\sigma_{jit_0}}{1+\sigma_{jit, min}/\sigma_{jit_0}}\right)$ and to adopt physically motivated lower and upper limits for $\sigma_{jit, min}$ and $\sigma_{jit, max}$, respectively \citep{Ford07}.

While information about the period and velocity amplitude comes exclusively from the RV observations, stellar jitter is correlated with chromospheric activity, so photometric observations or alternative analyses of spectroscopic observations could provide information about $\sigma_{jit}$ that is independent from the measured velocities.  
Often, observers may have some idea of what to expect for the stellar jitter, prior to analyzing the RV data, e.g., \citet{Wright05}.  
Of course, the stellar astrophysics is only one of multiple possible contributors towards the measured jitter.  
Either unmodelled instrumental noise and/or undetected planets may also cause excess RV scatter that is best modeled as a larger jitter given the presently available observations.   
Finally, we find it is quite useful to use a broad prior for $\sigma_{jit}$, especially during the initial exploratory phase, since it accelerates overcoming local minima by acting like  simulated annealing \citep{Ford06}.  
Therefore, we recommend adopting a broad prior for $\sigma_{jit}$ for initial analysis even when an estimate of the astrophysical jitter is available.  
If desired, the posterior can updated to incorporate any independent information about $\sigma_{jit}$ by adding an additional term to the likelihood and updating the posterior sample via either reweighting or rejection sampling \citep{Ford06}.

\section{Conventional MCMC for Posterior Sampling}
\label{secMcmc}
We described our likelihood function, $p(\vec{d}|\vec{\theta})$ in \S\ref{secL} and priors, $p(\vec{\theta})$, in \S\ref{secPrior}.  
We combine these in a Bayesian approach to calculate the posterior probability distribution for the model parameters given the observed data, $p(\vec{\theta}|\vec{d})$ using by Bayes' theorem,
\begin{equation}
p(\vec{\theta}|\vec{d})=\frac{p(\vec{d}|\vec{\theta})p(\vec{\theta})}{p(\vec{d})}.
\end{equation}
The evidence term $p(\vec{d})=\int_\theta p(\vec{d}|\vec{\theta})p(\vec{\theta})d\vec{\theta}$ is important when comparing different models but can be extremely difficult to compute especially in a high-dimensional model. 
A popular method for characterizing $p(\vec{\theta}|\vec{d})$ is a sampling algorithm known as Markov chain Monte Carlo (MCMC). 
The MCMC routine generates a sequence of {\em states}. Each state is a set of parameter values ($\theta$). 
Repeating the sampling procedure yields a {\em chain} of states that can be used to approximate the posterior probability distribution. 
Traditional summary statistics (e.g. mean, standard deviation) can be calculated from the posterior sample.
MCMC is a now a standard method in the astronomy community and has been applied to an array of astronomical data sets and problems \citep{Ford05}.
There are a number MCMC algorithms applied to RV datasets in particular (ExoFit \citep{Balan09}; HMCMC \citep{Gregory10}; emcee, \citep{Foreman-Mackey13}; EXOFAST \citep{Eastman13}) but most have employed a Keplerian model.

\subsection{Metropolis-Hastings Algorithm}
\label{secMH}
The most common sampling technique for MCMC involves a proposal distribution and the Metropolis-Hastings (MH) algorithm for deciding whether to accept or reject the proposal. 
A proposal state, $\vec{x}^{\prime}_{n}$, is generated using the parameters of the current state, $\vec{x}_n$. If the proposal distribution is symmetric, the MH acceptance probability reduces to
\begin{equation}
\frac{p\left(\vec{x}^{\prime}_{n}|\vec{d}\right)}{p\left(\vec{x}_n|\vec{d}\right)} \sim \exp\left[-\left(\chi^2_{eff}(\vec{x}^{\prime}_n)-\chi^2_{eff}(\vec{x}_n)\right)/2\right]
\end{equation}
If a random number drawn uniformly between 0 and 1 does not exceed $p\left(\vec{x}^{\prime}_{n}|\vec{d}\right)/p\left(\vec{x}_n|\vec{d}\right)$, then $\vec{x}^{\prime}_n$ is accepted as the new state of the Markov chain, $\vec{x}_{n+1}$; otherwise, it is rejected and $\vec{x}_{n+1} = \vec{x}_n$. 
Because the parameterization of both models are the same, $p(\vec{d})$ cancels out of the ratio and therefore does not need to be calculated in the MH process. 

\subsection{Random Walk Metropolis-Hastings MCMC}
\label{secRwMcmc}
Perhaps the most common proposal distribution is based on perturbing the model parameters from the present state, i.e., a random walk MH MCMC.  
The major drawback of the random walk MH algorithm is that its efficiency at drawing new states is strongly dependent on the direction and magnitude of the proposal distribution.  
If the algorithm changes the entire set of $\theta$ at once, then if even one parameter is perturbed too much, then trial states will rarely be accepted.  
This forces one to adopt a small scale for the size of the perturbations, which dramatically increases the time required for a Markov chain to traverse the target distribution.  
An alternative method is to implement a Gibbs sampler, which proposes new states changing only subset of $\theta$ while keeping the rest of the parameters fixed. 
This algorithm struggles when there are correlated parameters. 
In such a case, the Gibbs sampler has a difficult time traversing the posterior while an algorithm that could propose trial states in the principal direction of correlated parameters would more easily traverse across the target probability distribution. 

In principle, one can develop an intuition for reasonable proposal distributions.
For example, through a combination of physical intuition and trial and error, \citet{Ford06} identified variable transformations that resulted in efficient convergence of MCMC for RV observations.  
However, identifying efficient proposal distributions can be a time consuming process and is impractical for complex high-dimensional models.  
In principle, this can be automated, e.g., \citet{Ford06} and \cite{ Gregory11}, however these often result in dramatically increasing the computational cost relative to if a good proposal distribution were available.  
These issues become more important for high dimensional parameter spaces, as is necessary for modeling systems with multiple planets.
The improvement in the sampling efficiency becomes quite substantial for either systems with several planets or systems with strong interactions.
This can be easily understood, since increasing the number of planets and the strength of dynamical interactions both increase the correlation between model parameters, and it is the correlation between model parameters that results in poor performance of random walk MCMC methods.  

\section{Differential Evolution MCMC}
\label{secDemcmc}
\begin{figure}[!thb]
\begin{center}
\centerline{\includegraphics[scale=0.3]{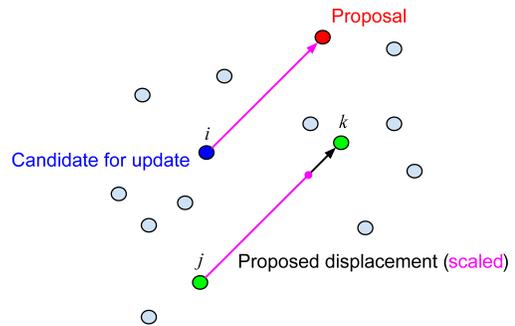}}
\caption{\small An illustration of the DEMCMC process. For a trial step for state $i$ (blue), we consider two additional states, $j$ and $k$ (both green). A proposal vector is drawn between $j$ and $k$, and the vector is scaled by $\gamma$ (magenta) before being added to state $i$ to generate a trial state (red).
\label{figDemcmc}}
\end{center}
\end{figure}

Our method of surmounting the above challenges is to replace the random-walk proposal distribution with a ``differential evolution'' proposal algorithm \citep{terBraak06}. 
Rather than using Gibbs sampling or picking a correlation structure for the proposal distribution, our DEMCMC algorithm considers a population of states within one ``generation.'' 
To evolve one generation of states to the next, DEMCMC creates a displacement vector between two randomly chosen states, $j$ and $k$. 
It then adds this vector to one state, $i$, from the current generation to generate a new trial state. 
This is mathematically expressed as
\begin{equation}
\vec{x}^{\prime}_{n,i}=\vec{x}_{n,i}+\left(\vec{x}_{n,j}-\vec{x}_{n,k}\right) \gamma.
\label{vectorEquation}
\end{equation}
where $\gamma=\gamma_0\left[1+z\right]$. 
Initially, we set $\gamma_0=2.38/\sqrt{2n_{dim}}$ (a scaling factor recommended by \citet{terBraak06}) and adjust $\gamma_0$ to adhere to a desirable acceptance rate (see \S\ref{secGamma}).  
Here $z$ is a random variable drawn from a Gaussian distribution with standard deviation $\sigma_\gamma$, a parameter to be discussed in \S\ref{secGamma}. 
Using the DEMCMC algorithm, such proposal steps naturally adapt their direction and scale based on the population of states in the current generation. 
For example, consider a population of states that are highly correlated. 
A random vector drawn between states roughly parallel to the principal correlation direction that results in steps being accepted is more likely to be large. 
Any vector drawn that points away (perpendicular) from the principal correlation direction will have a naturally small scale length. 
Figure \ref{figDemcmc} helps visualize the idea.

For the RV model parameters, we employ the following transformations for taking MCMC steps: $\log\left(1+P/P_0\right)$, $K\cos\left(\omega+M\right)$, $K\sin\left(\omega+M\right)$, $e\cos\omega$, $e\sin\omega$, $\Omega$, $i$, and $\log(1+\sigma_{jit}/\sigma_{jit_0})$ where $\sigma_{jit_0}=1$m/s based on recommendations of \citet{Ford06} for weakly interacting planetary systems.  

\subsection{Input Algorithm Parameters}
\label{secDemcmcInputs}
We implement the DEMCMC algorithm coupled to an N-body integrator for the self-consistent modeling of planetary systems and RV observations.  
The code is primarily C++ and is parallelized using either OpenMP (for multi-core workstations) or the Swarm-NG library for N-body integration using graphical processing units, GPUs \citep{Dindar13}.  
Henceforth we refer to the code as Radial velocity Using N-body DEMCMC (RUN DMC).  
This algorithm has been applied to several real RV datasets (HD 108874 \citep{veras10}, HD 200964 \citep{Johnson11}, HD 82943 \citep{Tan13}, 55 Cnc (B. Nelson et al. in prep.), GJ 876 (B. Nelson et al. in prep.)), but a thorough analysis of how RUN DMC performed when considering some key variables had not been done.
A number of input parameters need to be specified, including both algorithmic parameters (e.g., number of states per generation) and initial conditions to be used for the properties of the planetary system being modeled (e.g. number of planets and initial guesses for planetary masses and orbits).  
This section addresses the algorithmic parameters that may affect RUN DMC's performance.

\subsubsection{States Per Generation, $n_{chains}$}
\label{secStatesPerGen}
The number of states per generation is the size of our ensemble, analogous to the number of Markov chains being computed in parallel. In \S\ref{secPerform}, we explore how changing the value of $n_{chains}$ affects the performance of RUN DMC. 
First, we consider caveats for extreme values of $n_{chains}$. 
For $n_{chains}\le{n_{dim}}$, $n_{chains}$ points will sample the marginal posterior distribution projected onto an $n_{chains}-1$ dimensional subspace. 
The differential evolution aspect would prevent the chains from leaving this hyper-plane of the full parameter space. 
Therefore, we impose a hard lower bound of $n_{chains}>n_{dim}$. 
If $n_{chains}$ is not much larger than $n_{dim}$, then the states in a given generation may not give an accurate estimate of the covariance structure of the target distribution (since the number of possible proposal vector combinations is $n_{chains}\times[n_{chains}-1]/2$), leading to inefficient proposals and inefficient exploration of parameter space. 
However, increasing $n_{chains}$ also increases the number of computations per generation.
Thus, there exists a direct tradeoff between $n_{chains}$ and the number of generations ($n_{gen}$), since the total number of model evaluations is $n_{eval}=n_{gen}\times n_{chains}$. 
It is unclear whether it is advantageous for simulations to have more states per generation and fewer generations or fewer states per generation and more generations.  
We anticipate the optimal value for $n_{chains}$ will depend on $n_{dim}$. 
Thus, we test the performance of RUN DMC on simulated datasets with 1-4 planets in \S\ref{secPerturb}.

\subsubsection{Vector Scaling, $\gamma$ and $\sigma_\gamma$}
\label{secGamma}
As described in \S\ref{secDemcmc}, the trial states are generated by adding a scaled perturbation vector, $\gamma$, to the current state. 
We scale our proposal vectors by a factor $\gamma_0\left[1+z\right]$, where $z\sim N(0, \sigma_\gamma)$. If $\sigma_\gamma=0$, then $\gamma$ is constant, and the allowed proposal vectors will fall on a lattice, rather than filling parameter space, thereby hindering parameter space exploration. 
By setting $\sigma_\gamma>0$, RUN DMC is no longer restricted to sampling from a lattice. 
We consider values of $\sigma_\gamma$ across several orders of magnitude ($\sim$10$^{-4}$ to several $10^{-1}$) as a starting point. 
If we set $\gamma=1$, $\sigma_\gamma$ is the standard deviation of a Gaussian distribution centered on the tip of the proposal vector (Equation \ref{vectorEquation}). 
Since values of $\gamma$ close to 1 can be useful for multimodal posterior distributions, it can be advantageous to occasionally choose $\gamma$ near unity. 
However, most trial states should use a smaller value of $\gamma$ to achieve a desirable acceptance ratio. 
Therefore, we arbitrarily set $\gamma$ to 1 every 100 generations.

The value of $\gamma$ can be updated after every generation throughout a RUN DMC simulation. 
We aim for an acceptance fraction of 0.25. If too few states are being accepted ($<0.2$), $\gamma$ is scaled by 0.9 in the hope that smaller jumps will lead to a higher acceptance fraction. 
If the acceptance fraction exceeds 0.31, then $\gamma$ is scaled by 1.1 to allow for larger jumps. 
For intermediate acceptance fractions, $\gamma$ is scaled by $\sqrt{\textup{Acceptance Fraction}/0.25}$. 
In DEMCMC, this procedure references information from only one previous generation of states, so our algorithm is still Markov for each generation. 
The mathematical conditions for RUN DMC converging to the target distribution are still satisfied, and thus, adjustments in the proposal vector size will not change the shape or scale of the target distribution.

\subsection{Choosing States for the Initial Generation}
\label{secInitGen}
DEMCMC is most efficient when the initial ensemble of states is close to a posterior sample. 
For RV datasets, a Keplerian model usually provides a good first approximation to the RV curve that would be calculated from an N-body integration.  
Thus, the posterior sample from an analysis based on a planetary model using a linear superposition of planets on independent Keplerian orbits can be used with standard techniques for a short MCMC simulation.
A posterior sample from the much faster Keplerian MCMC \citep{Ford06} simulation can be used for the states in the initial generation of the DEMCMC simulation.  
Since the MCMC output is not used for inference, but rather as the starting point for the DEMCMC, it is not necessary for the MCMC sample to pass extensive convergence tests.
Since the Keplerian model is insensitive to orbital inclination and the longitude of ascending node, we assign these randomly, drawing from their prior distributions after imposing any constraints.  
For example, this paper focuses on coplanar systems, so we assign inclinations of $0<\cos(i)<1$ and $\Omega$ of 0$^\circ$.  
Note that in \S\ref{secPerform}, we want to test how well the DEMCMC algorithm recovers from an initial ensemble that is not close to a posterior sample.
Therefore, we will intentionally perturb the initial conditions so that they are not a good approximation to the posterior (see \S\ref{secPerturb}).  

\subsubsection{Adaptive Target Distribution, MassScaleFactor}
\label{secMassScaleFactor}
What if the dynamical interactions of a particular system are so strong that the Keplerian and N-body solutions occupy different regions of parameter space with little or no overlap? 
We can gradually change the target distribution from one very close to the posterior for the Keplerian model to the posterior for the N-body model. 
To achieve this, we utilize the MassScaleFactor parameter, which allows us to slowly ``turn on'' the N-body effects. 
For example, consider a transformed problem where we multiply the planet masses, the measured RVs, and their respective observational uncertainties by MassScaleFactor=0.001. 
A system originally with two Jupiter-mass planets (that could exhibit strong orbital interactions if near a MMR) is transformed to a system with two sub-Earth mass planets in nearly the same orbital configuration. 
For such small planetary masses, the Keplerian and N-body solutions would be indistinguishable for a typical set of RV observations, so the posterior distributions will have significant overlap. 
For such a DEMCMC simulation, we would start with MassScaleFactor=0.001 and gradually increase the value of MassScaleFactor with each generation until MassScaleFactor=1.0 for generations greater than $0.1\times{n_{gen}}$, so the target distribution approaches the desired posterior for the N-body model \citep{Laughlin01b}. 
No states from generations prior to $0.1\times{n_{gen}}$ are used for the inference, so the use of MassScaleFactor does not offset the posterior sample, except to help accelerate convergence of the RUN DMC algorithm in the face of a poor initial ensemble due to strong N-body interactions.
Since this study focuses on weakly interacting planetary systems, we set MassScaleFactor to unity throughout the DEMCMC simulation unless otherwise noted.  

\section{Algorithm Performance}
\label{secPerform}

Traditional, random walk MCMC algorithms often struggle to sample efficiently from high dimensional parameter spaces, such as those required for modeling a multiple planet system.  We will show that the DEMCMC algorithm excels in navigating the parameter spaces for planetary systems with several planets.
We performed a series of tests to see how well RUN DMC performed for various datasets and planetary system models.
There are two aspects to RUN DMC's performance we want to test: the duration of the burn-in phase required (i.e. how efficiently the algorithm can recover from an inaccurate initial guess for the posterior distribution) and the efficiency of the sampling algorithm once the population of states are all near a single (presumably global) posterior mode.
We are particularly interested in evaluating how well the DEMCMC algorithm samples from the posterior distribution for planetary systems with three or more planets.

\subsection{Method for Perturbing Initial Ensemble}
\label{secPerturb}
Our first series of tests were designed to determine how well RUN DMC could recover from an inaccurate initial population of states. 
We generated synthetic datasets using the masses and orbital properties based on real exoplanet systems maintaining observation times and uncertainties of the actual RV time series. 
We considered datasets based on the following systems: HIP 75458 (one planet), HD 12661 (two planets), and HIP 14810 (three planets). 
By design, these systems have negligible dynamical effects for their respective observing baselines, so we are able to explore the effects of increasing the number of planets, and thus the dimensionality of the parameter space to be explored, without worrying about the strength of mutual planetary interactions.   
We will explore the efficiency of DEMCMC for strongly interacting systems in \S\ref{secAutoCorDyn}.  
We also consider the RV datasets of $\mu$ Ara (four planets) and 55 Cnc (a four-planet version excluding the inner-most planet, 55 Cnc e). 
The primary objective of these tests is to gauge how the performance of the ``differential evolution'' aspect of RUN DMC is affected by $n_p$ and the algorithmic parameters (e.g. $n_{chains}$, $\sigma_\gamma$).

Strictly speaking, our results are most applicable to studies of these specific planetary systems, since the shape of the posterior distribution, and thus the complexity of sampling from it, depends on the details of both the planetary system being observed and the properties of the observational data.
Nevertheless, our detailed investigations of the performance of DEMCMC for these planetary systems can provide valuable insight into the algorithm's anticipated performance for other radial velocity data sets.
Algorithms or parameter values that do not perform acceptably for these test cases are unlikely to be worth applying to similar planetary systems.
Similarly, choices of algorithmic parameters that result in desirable performance for these test cases can be adopted as the initial parameters for analysis of other planetary systems, or even these same planetary systems as additional data become available.

We begin each test with a posterior sample of the masses and orbital parameters obtained from a fully converged MCMC sample based on a Keplerian model. 
In order to simulate a larger disparity between the initial population and the target distribution, we generate an initial population by perturbing a subset of the states from the converged MCMC sample in one of three ways. 
For the first perturbation method, we {\em scattered} the states by increasing the dispersion about their median values, i.e. if a particular parameter distribution is well defined by a Gaussian with standard deviation 1-$\sigma$, a scatter of $\alpha$ would produce a Gaussian distribution with a dispersion of $\alpha$-$\sigma$. 
If we consider discrete values in our set of model parameters, $\vec{x}_i$ for $i=1$ to $n_{chains}$, and determine the median value $\left \langle \vec{x} \right \rangle$ of each parameter, then the scattered values, $\vec{x}_\alpha$, are calculated as follows:
\begin{equation}
\vec{x}_{i, \alpha}=\left \langle \vec{x} \right \rangle + \alpha\left(\vec{x}_{i} -  \left \langle \vec{x} \right \rangle \right).
\end{equation}

For the second perturbation method, we apply a global $shift$, or displacement, of the states in parameter space, i.e. a shift with a scale factor of $\beta$-sigma corresponds to increasing every parameter value by that $\beta$ for the input population, where $\vec{\sigma}$ is the vector of standard deviations of the parameter contained in $\vec{x}$. 
Mathematically speaking,
\begin{equation}
\vec{x}_{i, \beta}=\vec{x}_{i} + \beta\vec{\sigma}
\end{equation}

To avoid possible boundary condition issues with eccentricity and angles, we perturbed $e\sin\omega$, $e\cos\omega$, and $\omega+M$ rather than $e$, $\omega$, and $M$. 
The angles are constrained to the domain of 0 to $2\pi$ radians. 
Thus for the purpose of generating initial conditions with a shift or scatter, $\vec{x}_{i}=\left\{P_1, K_1, e_1\sin\omega_1, e_1\cos\omega_1, \omega_1+M_1, \ldots\right\}$.

We also considered a third perturbation that combined $\alpha$ and $\beta$, i.e.
\begin{equation}
\vec{x}_{i, \alpha\beta}^{\prime}=\left \langle \vec{x} \right \rangle + \alpha\left(\vec{x}_{i} -  \left \langle \vec{x} \right \rangle \right) + \beta\sigma_{i},
\end{equation}
to see if a scatter of the initial shifted ensemble could help shorten the burn-in phase. In summary, we performed a $\beta=3$ followed by an $\alpha=3$ perturbance on an ensemble of states for the two and three-planet cases. For the two-planet case, the effect of the scatter was negligible compared to just a $\beta=3$ perturbance. For the three-planet case, the scatter resulted in an extended burn-in, by roughly 100 generations across all $n_{chains}$ and $\sigma_\gamma$. We conclude that scattering a shifted ensemble is not likely to accelerate convergence.

\subsection{Method for Estimating Required Number of Generations for Burn-in Phase}
\label{secPerformAlg}

\begin{figure}[!thb]
\begin{center}
\centerline{\includegraphics[scale=0.4]{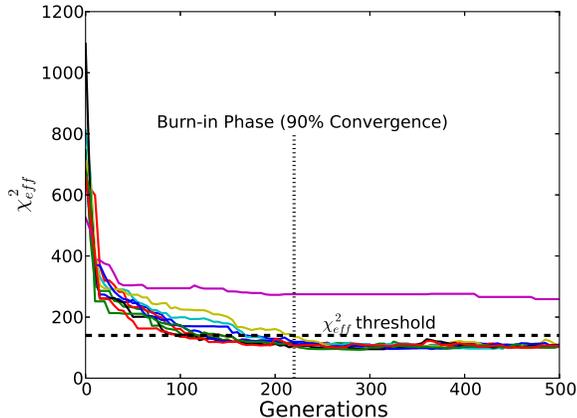}}
\caption{\small $\chi^2_{eff}$ as a function of generation in an example of a RUN DMC simulation with ten chains displayed. Within 220 generations, nine of the chains recovered to the mode of the global minimum while the one became trapped in a local minima. By monitoring and determining a threshold value for $\chi^2_{eff}$, we define the burn-in phase as the minimum generation for 90\% of chains to have $\chi^2_{eff}$ smaller than the threshold value. 
\label{figChiSq}}
\end{center}
\end{figure}

For each of the synthetic systems (with $n_p=\left\{1, 2, 3, 4\right\}$), we performed a series of RUN DMC simulations for a variety of $n_{chains}$ and $\sigma_\gamma$ values. 
We fixed $n_{eval}=512,000$ across all values of $n_{chains}$, in order to determine which parameter ($n_{chains}$, $n_{gen}$) had a greater effect on the convergence rate. 
For example, a simulation done with 16 $n_{chains}$ ($n_{gen}=32,000$) ran for 4 times as many generations as one with 64 $n_{chains}$ ($n_{gen}=8,000$). 
For every value of $n_{chains}$, we computed the required number of generations for the burn-in phase, which we define as the number of generations before which 90\% of the states within the final generation have a $\chi_{eff}^2$ less than an experimentally obtained threshold value (see Figure \ref{figChiSq}). 
We performed 2048/$n_{chains}$ RUN DMC simulations for each value of $n_{chains}$ (16, 32, 64, 128, 256, 512, 1024) and for each value of $\sigma_\gamma$ (0.0001, 0.0016, 0.0256, 0.4096). 

\subsection{Results for Testing Converged Chains}
\label{secConvergence}
We considered a range of values for $\alpha$ and $\beta$ for different planetary systems. 
Tests with a small to significant scatter ($0.1<\alpha<10.0$) or shift ($0.1<\beta<1.0$) typically lead to 100\% of the chains recovering to the mode of the global minimum by the end of their respective simulations.
As expected, very large perturbations (e.g., $\beta\gtrsim40, 20, 15$ for the two, three and four-planet cases, respectively) lead to a small fraction of chains finding the global minimum.
Due to the symmetrical nature of a scatter perturbation, some states are likely to reside near the global mode of the posterior distribution.  
We find this occurs even for the high-dimensional parameter spaces, such as when we consider a four-planet model.

Finding the global minimum after a $\beta$ perturbation is sensitive to $n_{dim}$ and a longer burn-in period is often required for DEMCMC simulations to converge on the global mode. 
The rest of \S\ref{secConvergence} is dedicated to explaining the results of these tests, which have been condensed into Figure \ref{figConvergeAll}.
There are a few caveats for interpreting this figure. 
First, $n_{chains}=16$ is excluded, since this value provided generally poor results, as it is comparable to or even less than $n_{dim}$ for the synthetic systems simulated.
We want to easily visualize potential correlation structures for the parameters that affect performance.
Second, one must draw attention primarily to the different scalings of the $y$-axis panels to properly see the correlation structure and how it varies with $n_{chains}$, $\sigma_\gamma$, and $\beta$.

Lastly, one should consider the typical differences in the RV datasets as a function of the number of planets.
Observers typically collect more RV observations prior to claiming a planet detection and publishing observations for systems with more known planets.
While, we would intuitively expect a general increasing trend in the length of the burn-in phase as a function of $n_p$ at a fixed $\beta$, the challenge of the increased dimensionality is often partially offset by the additional observational information.  
This makes it difficult to compare directly to results for systems with different number of planets.
Ultimately, we are interested in analyzing real data sets.
Therefore, we have opted to use simulated datasets based on the actual number, observing times, and estimated measurement uncertainties for a known planetary system.

Strictly speaking, the above results are specific to the planetary systems and data sets being investigated.  Nevertheless, this figure provides guidance for what to expect when analyzing typical radial velocity data sets as a function of the number of planets.  We caution that particular care is necessary when analyzing strongly interacting systems, as we will discuss in more detail in \S\ref{secAutoCorDyn}.

\subsubsection{One Planet}

First, we consider a single planet system using simulated data based on observations of HIP 75458, a K star harboring a super-Jupiter-mass planet on an eccentric 510 day orbit. 
The actual data has a long-term RV trend suggestive of a second companion on a long-period orbit. 
We base our simulated observation times and measurement uncertainties on the actual observation times and uncertainties and generate synthetic data using a single planet model based on the best-fit model parameters from \citet{Butler06}.

In the first scattering ($\alpha$) test, we found that RUN DMC rapidly recovers ($n_{gen}<100$) from an initial perturbation of $\alpha=\{0.25, 2, 5\}$ with a near-unanimous ($>0.98$) convergence across all values of $n_{chains}$ and $\sigma_\gamma$ by the end of each simulation. 
For larger values of $n_{chains}$, we found a slightly reduced fraction of chains converging. 
Presumably, this is due to the reduced number of generations since we held the $n_{eval}$ fixed. 
Even more highly scattered initial conditions ($\alpha>10$) resulted in a smaller fraction of converged chains, across all values of $n_{chains}$, but still performed well ($>0.95$ convergence by the end).

In the second test shifting ($\beta$), we observed similar behavior. 
Larger offsets of the initial population resulted in a lower convergence fraction, although for every case tested ($\beta=\{1,3,5\}$), over 90$\%$ of chains had recovered by the final generation. 
Overall, one-planet systems seem to be fairly insensitive to $n_{chains}$ across a few orders of magnitude, but for the best performance, our recommendation is to use $16<n_{chains}<64$ with relatively long chain lengths. 
Note that 32,000 generations is still much less than is often needed for standard MCMC analyses, e.g., \citet{Ford06}.

\subsubsection{Two Planets}

Next, we consider a two-planet system using simulated data based on observations of HD 12661, a quiet G-main sequence star with two known 2-Jupiter mass planets. 
The inner planet has a moderately eccentric, 262 day orbit. 
The outer planet has a 1700 day orbital period and has undergone a couple of orbits in our observing baseline. 
We base our simulated observation times and uncertainties on the actual observation times and uncertainties and generate synthetic data using planet masses and orbital parameters based on the best-fit model parameters from \citet{Butler06}.

\begin{figure*}[!htb]
\begin{center}
\centerline{\includegraphics[scale=0.45]{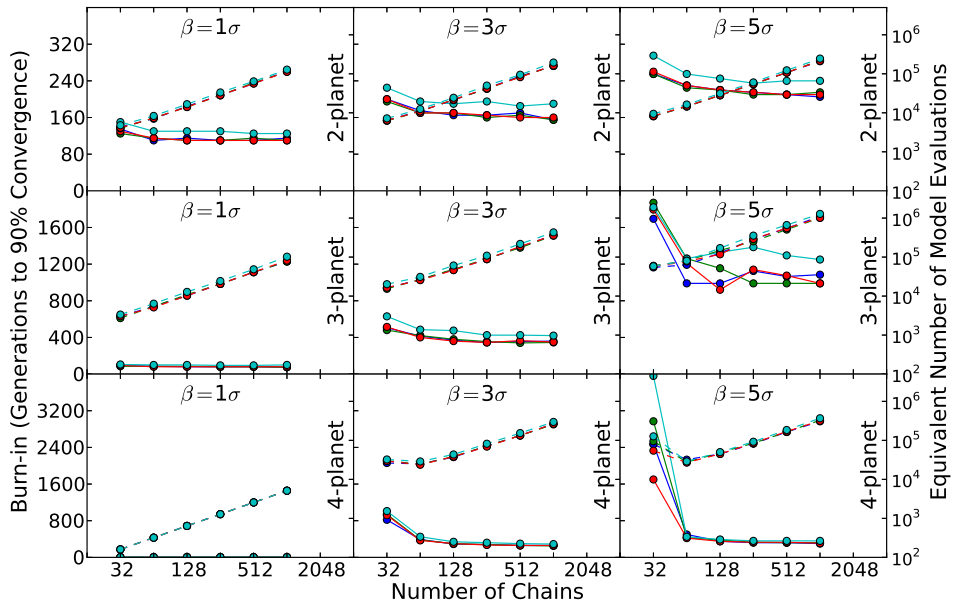}}
\caption{ \small Each panel shows the burn-in time for 90\% of chains to converge to the posterior mode (vertical axis) as a function of $n_{chains}$ (horizontal axis) for various values of $n_p$, $\beta$, and $\sigma_\gamma$. Each vertical column represents a fixed value for $\beta$ (left column, 1; middle column, 3; right column, 5), and each horizontal row represents a synthetic $n_p$ planet system (top, two-planets; middle, three-planets; bottom, four-planets [large dataset, 55 Cnc]) based on datasets and orbital properties mentioned in \S\ref{secConvergence}. Each colored line shows results of a simulation with a different value for $\sigma_\gamma$ (blue=0.0001; green=0.0016; red=0.0256; cyan=0.4096). These can be seen as a function of number of burn-in generations (left vertical axis scale, solid lines) or as a function of number of total model evaluations (right vertical axis scale, dashed lines). We find a slight preference toward large values of $n_{chains}$ for reduced the required number of generations of burn-in. 
\label{figConvergeAll}}
\end{center}
\end{figure*}

For an unreasonably large value of $\alpha=10$, RUN DMC requires a few 100s of generation to achieve 90\% convergence.
However, the $\beta$-type perturbation results were quite different (Figure \ref{figConvergeAll}, top three panels). 
We considered $\beta=\{1, 3, 5, 20, 40, 80\}$; of these, $\beta=40$ was the largest value that resulted in a non-negligible fraction of chains succeeding in finding the global mode. 
For such an unrealistically large $\beta$, only a small fraction of chains reached the global mode of parameter space for extreme values of $n_{chains}$ (16, 1024). 
This was most likely due to $n_{chains}$ not being significantly larger than $n_{dim}$ (see \S\ref{secStatesPerGen}) and the smaller $n_{gen}$, respectively. 
Perhaps, more remarkable is that RUN DMC was able to find the global posterior mode within a reasonable number of generations for any $n_{chains}$.  
Fortunately, this is much greater than any the offset between Keplerian and N-body solution for any known two-planet systems. 

The burn-in phase lasts approximately 100 generations for $\beta=1$, and this value scales up in a roughly linear fashion with increasing $\beta$ within this limit of realistic $\beta$ offsets. 
For the extreme value of $n_{chains}=16$, the burn-in phase lasted over 1,000 generations for $\beta=\{1, 3, 5\}$.
We left these data values out of Figure \ref{figConvergeAll} in order to see the general trends with $n_{chains}$ and $\beta$. 
Our results seem to be insensitive to $\sigma_\gamma$ across many orders of magnitude, though our largest value (0.4096) was slightly less efficient. 
We speculate that a large $\sigma_\gamma$ causes the proposal distribution scale ($\gamma_0$) to vary so much that it does not settle on a value with a desirable acceptance rate, slowing convergence of the Markov chains.  

Given a choice between the two, it is generally preferable to choose larger $n_{gen}$ rather than larger $n_{chains}$ (provided $n_{chains}$ is significantly larger than $n_{dim}$; \S\ref{secStatesPerGen}) for computer architectures where run time scales linearly with the number of model evaluations.  
For parallel architectures where performance increases with the number of states per generation (e.g., GPUs, cloud), a larger $n_{chains}$ may result in a smaller wall clock time. 
The dependence of the rate of convergence on the chosen $\sigma_\gamma$ is weak across many orders of magnitude.

\subsubsection{Three Planets}
Next, we test RUN DMC on a simulated dataset of a three-planet system. 
HIP 14810 is a G star harboring a three-planet system composed of Jupiter mass bodies in a widely spaced, moderately eccentric configuration. 
We base our simulated observation times and uncertainties on the actual observation times and uncertainties and generate synthetic data using planet masses and orbits based on the best-fit parameters from \citet{Wright09b}.

For this test three-planet system and $\alpha=10$, we observe a sharp decrease in the fraction of converged chains if $n_{chains}\leq32$. 
This can be attributed to a relatively large $n_{dim}$. 
We see a similar trend as in the two-planet case, except the burn-in phase has scaled up significantly. 
The dependence on $\beta$ is also moderately stronger, but the effects of varying $\sigma_\gamma$ are still weak across many orders of magnitude.

For this particular three-planet system, $\beta=20$ was the largest value to succeed in finding the global mode, still much greater than the typical discrepancies between Keplerian and N-body solutions (e.g., HD 200964 and 24 Sextanis \citep{Johnson11}, 55 Cnc (B. Nelson et al. in prep.)). 
Simulations with $n_{chains}=32$ required significantly more generations of burn in than simulations with $n_{chains}\gtrsim64$.
A three-planet system settles for a narrower range of $n_{chains}$ ($\sim$100) and $n_{gen}$ values ($\sim$10$^4$) for $\beta=20$. 
Simulations with $n_{chains}=512$ or 1024 resulted in a smaller fraction of chains reaching the global mode, likely due to the reduced number of generations.

\subsubsection{Four Planets}
\label{secFourPlanets}
We test RUN DMC using two four-planet systems. 
First, we consider 55 Cnc A, an aging, nearby G star with a five-planet system and one of the longest observed RV targets. 
55 Cnc A has a long period stellar companion, 55 Cnc B, which we do not account for in this analysis.
Previously, the system has been analyzed assuming a coplanar, Keplerian model. 
We base our simulated observation times and uncertainties on the actual observation times and uncertainties from Lick and Keck data \cite{Fischer08} and consider a four-planet model that excludes the inner-most low-mass planet (55 Cnc e). 
For 55 Cnc, we use a Keplerian model to generate simulated data and and analyze the simulated observations with MassScaleFactor=0.01 for the entire simulation to avoid any affect due to significant planet-planet interactions. 
This series of tests for a four-planet system is a precursor to a more thorough analysis of the dynamical architecture of the 55 Cnc system that will be presented in an forthcoming paper (B. Nelson et al. in prep.). 
The information from these tests will help us determine how to approach the 55 Cnc dataset once we include a much more computationally expensive self-consistent N-body model with all five planets. 

In Figure \ref{figConvergeAll}, we show the results for 55 Cnc in the bottom three panels as our example of a four-planet system. 
For $\beta=1$, all simulations found the global mode within just 5 generations, most likely due to the comparatively large number of observations and high signal-to-noise of most of the planets.  
Increasing $\beta$ leads to a longer burn-in as expected.
RUN DMC begins to suffer when using relatively small values of $n_{chains}$ (i.e. 32), presumably due to $n_{chains}$ being only slightly larger than $n_{dim}$. 
A simulation with $\beta=15$ had the largest $\beta$ value for which a non-negligible fraction of chains succeeded in reaching the global mode.  

Next, we consider the $\mu$ Ara planetary system, a naked eye G star that hosts four known planets. 
We base our simulated observation times and uncertainties on the actual observation times and uncertainties and generate synthetic data using planet masses and orbital parameters based on the best-fit parameters from \citet{Butler06}. 
We expect the $\mu$ Ara system to be more challenging than the 55 Cnc system, since the $\mu$ Ara planets induce an RV perturbation with a typically lower signal-to-noise than the 55 Cnc planets. 
Additionally, we consider 108 $\mu$ Ara RVs, significantly less than the 320 RVs for 55 Cnc.
For $\mu$ Ara, a $\beta=1$ perturbation required a burn-in of a couple hundred generations for chains to find the global mode. 
RUN DMC did not succeed in achieving 90\% convergence across all $n_{chains}$ for $\beta=\{3, 5\}$. 
For these values of $\beta$, roughly half the states lingered in low probability regions or local minima.
Thus, we investigate intermediate values of $\beta=\{1.5,2\}$.
Both struggle to converge for $n_{chains}=32$ (565-850 and 1360-2590 generations, respectively) but $n_{chains}\geq64$ performs significantly better (375-435 and 680-990 generations, respectively).
Much like previous results for two and three-planet systems, there is a gradual improvement in the burn-in phase with increasing $n_{chains}$.

To better understand what properties of the planetary system and observations were most responsible for the difference between the results for the two four-planet systems, we tested a case where we considered only the first 108 RVs of our 55 Cnc dataset (i.e., equal to the number of RVs used for $\mu$ Ara).  The RVs were chosen so that the resulting observing baseline for 55 Cnc was comparable to that of the $\mu$ Ara observations.
After obtaining a initial ensemble from this new dataset, we carried out multiple tests using $\beta$ perturbations with $\beta=\{1,3,5\}$.  
Simulations with $\beta=\{1,3\}$ converged within a few hundred generations.
However,  the majority of our ensemble for the $\beta=5$ simulation did not satisfy our convergence criterion. 
We conclude that for challenging systems with the same dimensionality, RUN DMC converges more rapidly (at least in terms of number of model evaluations) if a larger number of observations are available to provide strong constraints on the posterior distribution. 

\subsection{Autocorrelation Time for Various Systems}
\label{secAutoCor}
The previous tests for convergence allow us to estimate the length of the burn-in phase required before we can be confident in using states from the Markov chains for inference. 
Next, we ask ``how efficiently does each chain sample parameter space near this posterior mode after an adequate burn-in?'' 
We consider the behavior of RUN DMC once it has arrived near the global mode and completed burn-in. 
We quantify the efficiency of a RUN DMC simulation by calculating the autocorrelation (AC) of several model parameters as a function of the number of generations between two states of a Markov chain (referred to as the ``lag''). 
A small magnitude of the AC for a shorter lag implies that a Markov chain of a given length would contain more effectively uncorrelated states. 
Much like the previous convergence tests, we want to determine how the AC function of a certain parameter depends on $n_{chains}$ and $\sigma_\gamma$. 
More broadly, we investigate the dependence of AC on $n_{dim}$ and the strength of the dynamical self-interactions of planetary systems.

\begin{figure}[!htb]
\begin{center}
\centerline{\includegraphics[scale=0.4]{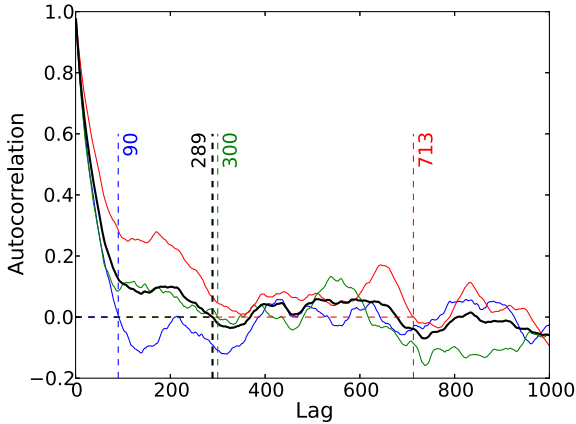}}
\caption{\small An example of the AC function for orbital period for the two-planet system as a function of generation lag. The red, green, and blue lines are AC functions of three different chains. The bold black line is the average of the three aforementioned curves. The vertical dashed lines indicate the minimum lag at which the AC$\leq$0, or AC$^{-1}(0)$. This metric will be the basis for the subsequent figures.
\label{figAcEx}}
\end{center}
\end{figure}

For robustly estimating the AC for a given $n_p$, $n_{chains}$, and $\sigma_\gamma$, we perform $n_{sim}$ simulations, so that $n_{sim}\times n_{chains} \geq 256$. 
We compute the autocorrelation as a function of lag of each chain and calculate the average AC value over all the chains as a function of lag. 
We show an example with three AC functions and one averaged AC function in Figure \ref{figAcEx}.
We estimate the minimum lag such that AC(lag) is negative and use the notation ``AC$^{-1}(0)$'' to represent this metric. 
Note that since the autocorrelation function is not invertible, AC$^{-1}(0)$ should not be taken literally to mean the inverse of the autocorrelation function evaluated at 0, but rather it denotes the concept of the approximate number of generations needed for effectively independent samples.  

\subsubsection{Results for Autocorrelation Length as a Function of $n_p$ or $n_{dim}$ }
\label{secAutoCorNpl}

\begin{figure*}[!htb]
\begin{center}
\centerline{\includegraphics[scale=0.45]{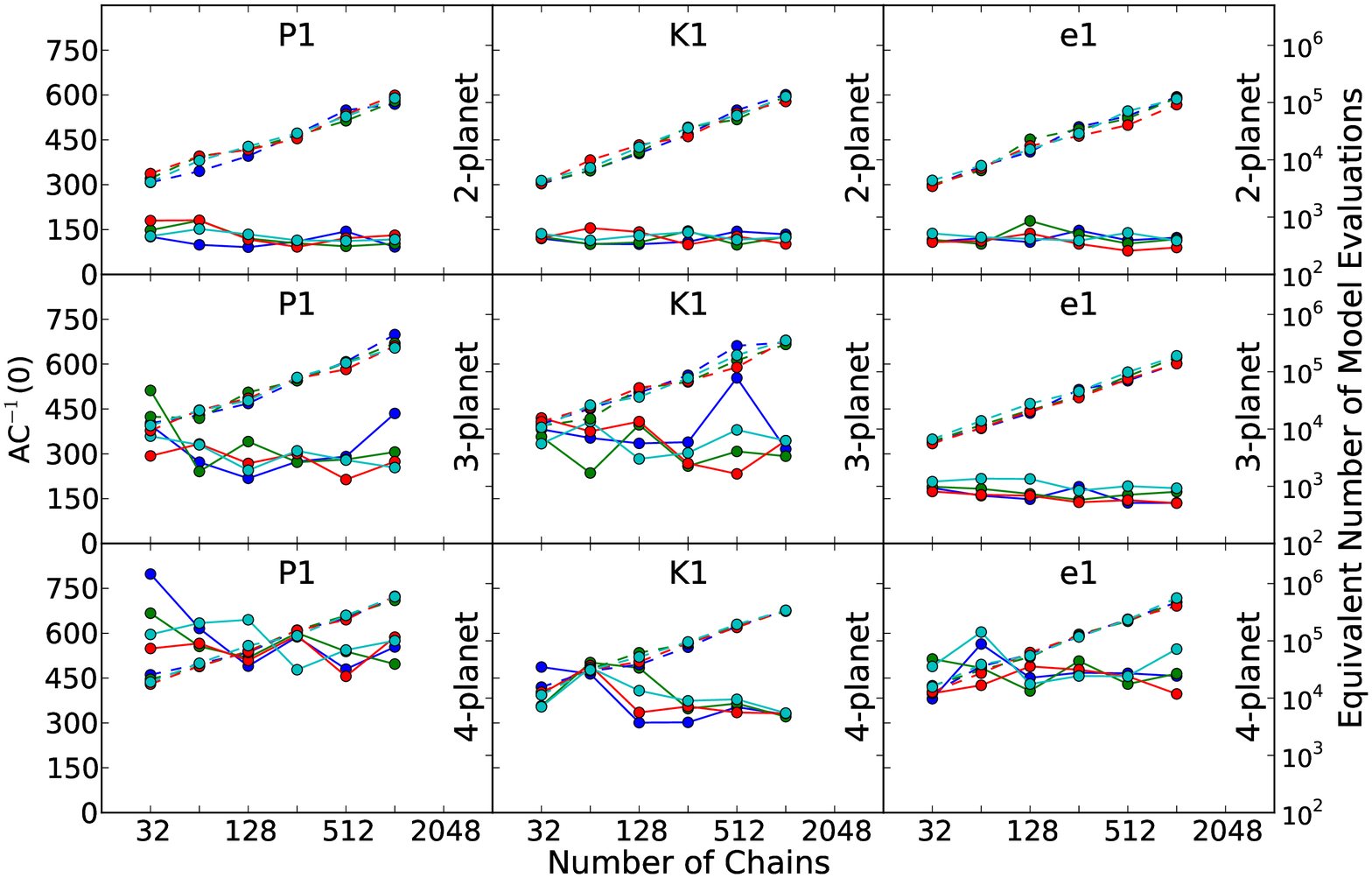}}
\caption{\small Each panel shows the minimum lag at which AC$\le$0 as a function of $n_{chains}$ and $\sigma_\gamma$ (colored lines) for one parameter and simulated dataset. Each horizontal row represents a synthetic $n_p$ planet system (top, two-planets; middle, three-planets; bottom, four-planets [$\mu$ Ara]) based on datasets and orbital properties mentioned in \S\ref{secConvergence}. We show results for three representative parameters for the inner-most planet in all of these simulations (left column, $P$; middle column, $K$; right column, $e$). 
\label{figAcAll}}
\end{center}
\end{figure*}

In Figure \ref{figAcAll}, we analyze the exact same systems and observations from \S\ref{secConvergence} (with $\mu$ Ara as the four-planet system). 
For these simulations, we start with an initial ensemble of states drawn from near the global mode based on a previous MCMC simulation based on a Keplerian model. 
Using the same set of values for $n_{chains}$ and $\sigma_\gamma$ as in Section 5.1, we set $n_{gen}=11,000$, and as a precaution, we treat the first $1,000$ generations as burn-in. 
Since $n_{gen}$ is fixed, runs with larger $n_{chains}$ require the most model evaluations and thus the longest time to complete. This should be kept in mind when interpreting Figures \ref{figAcAll}-\ref{figAcCompact}.

We see that simulations with larger $n_{chains}$ tend to result in slightly smaller AC$^{-1}(0)$ in Figure \ref{figAcAll}. 
We speculate that a larger population of states provides more proposal vector combinations, a better estimate of the covariance structure, and thus moderately more efficient sampling around the global mode. 
However, given the tradeoff between $n_{gen}$ and $n_{chains}$, we would only  recommend using large ($\sim$1000) values of $n_{chains} \gg n_{dim}$ when run on a highly parallel architecture (e.g., GPU or clouds). 
The differences amongst order of magnitude variations in $\sigma_\gamma$ were not significant. 
From two to three to four-planet systems, there is an overall significant increase the value of AC$^{-1}(0)$ primarily due to the increase in $n_{dim}$. 
The 55 Cnc system with additional RVs had slightly better AC$^{-1}(0)$ values than the three-planet case. 

\subsubsection{Results for Autocorrelation Length as a Function of Dynamical Strength}
\label{secAutoCorDyn}
\begin{figure*}[!htb]
\begin{center}
\centerline{\includegraphics[scale=0.45]{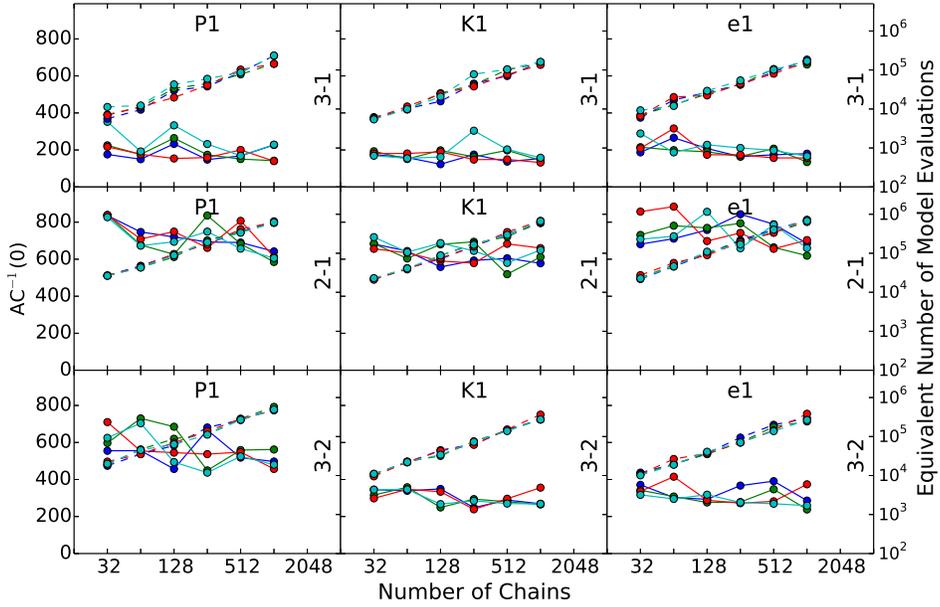}}
\caption{\small Each panel shows the minimum lag at which AC$\le$0 as a function of $n_{chains}$ and $\sigma_\gamma$ (colored lines) for each panel. Simulations were performed for synthetic, long-period two-planet systems near a mean-motion resonance: 3:1 (top row), 2:1 (middle row), and 3:2 (bottom row) systems based on datasets and orbital properties mentioned in \S \ref{secAutoCorDyn}. Here we show results for three representative parameters of the inner-most planet in each of these simulations (left column, $P$; middle column, $K$; right column, $e$).  
\label{figAcWide}}
\end{center}
\end{figure*}
So far, we have limited our analysis on planetary systems that are well approximated by Keplerian orbits in order to explore how the performance of RUN DMC depends on the number of model parameters.
For strongly interacting exoplanet systems, there may be more complex and non-linear parameter correlations. How is the algorithm performance affected by such interactions? 
For these tests, we created six synthetic two-planet systems each near one of three mean-motion resonances (3:1, 2:1, 3:2).
For each resonance, we create two simulated data sets: one labelled ``long-period'' (with orbital periods of hundreds of days) and one labelled ``short-period'' (with orbital periods of tens of days).
For both of these cases, we based our observation times and uncertainties on HIP 75458 \citep{Butler06}, a K giant harboring in eccentric gas giant and suspected to have another massive companion in a wide orbit. 
The time series itself has phase coverage that would be hypothetically consistent with observing patterns for either ``long-period'' or ``short-period'' systems. 
We construct systems of 1 $M_J$ and 2 $M_J$ planets orbiting a $1M_\odot$ star with orbital periods of either 400 and 1212 days, 400 and 808 days, or 400 and 606 days.
The question of whether or not a pair of planets orbit in a mean-motion resonance is extremely sensitive to the masses and orbital architectures, so for consistency, we set the period commensurability to a value slightly larger than 3, 2, and 1.5 respectively.
Long term integrations confirm that the resonant arguments for these systems are circulating, so none of these systems are technically in a mean-motion resonance. 
Thus, these systems are similar to the planetary systems discovered by Kepler in terms of their proximity to resonance \citep{veras12}.
In each system, the planets have $e_{inner}=0.04$ and $e_{outer}=0.02$ and the planets have pericenter directions that are initially apsidally aligned about $0^o$ ($\omega_b-\omega_c=0^o$). 
The short-period system consists of planets of the same masses, eccentricities, and orbital alignment but with the following orbital periods: 40 and 121.2 days, 40 and 80.8 days, and 40 and 60.6 days.
For the sake of simplicity, these systems will be referred to as 3:1, 2:1, and 3:2 configurations.

As mentioned \ref{secAutoCor}, we average the autocorrelation of 256 chains and perform multiple runs for RUN DMC simulations with less than 256 $n_{chains}$.
For each combination of $\sigma_\gamma$ (0.001, 0.0016, 0.0256, 0.4096) and orbital configuration (3:2 wide, 2:1 wide, 3:1 wide, 3:2 close, 2:1 close, 3:1 close), we perform 17 runs each for 11,000 generations. 
In total, Figures 6 and 7 summarize the results of 408 RUN DMC simulations.
As each model evaluation requires performing an n-body integration, this represents a significant amount of computation.
We estimate that these tests alone required $\sim$2300 CPU hours.

\begin{figure*}[!htb]
\begin{center}
\centerline{\includegraphics[scale=0.45]{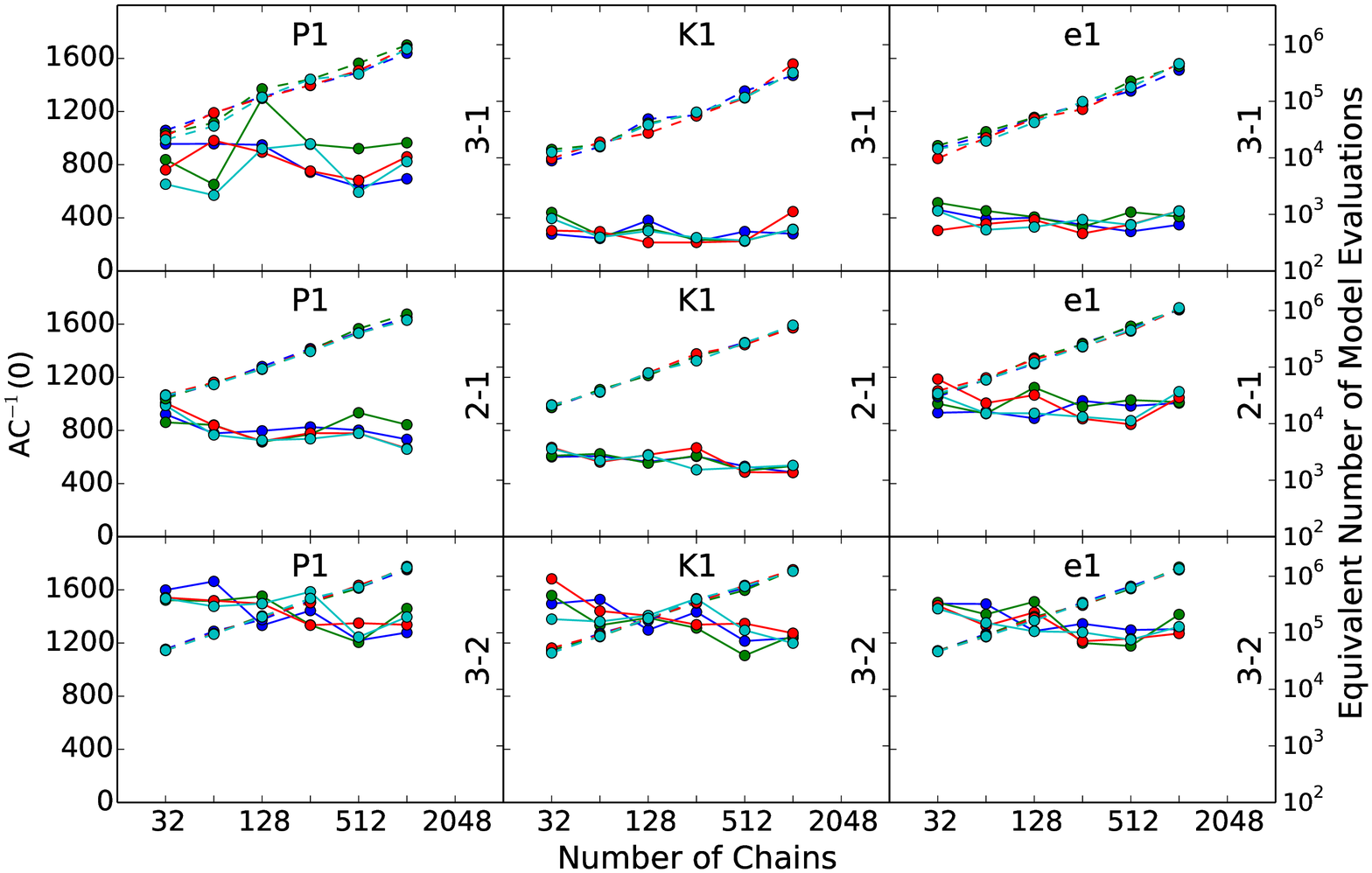}}
\caption{\small Each panel shows the minimum lag at which AC$\le$0 as a function of $n_{chains}$ and $\sigma_{\gamma}$ (colored lines) for each panel. Simulations were performed for synthetic, closely separated two-planet systems near a mean-motion resonance: 3:1 (top row), 2:1 (middle row), and 3:2 (bottom row) systems based on datasets and orbital properties mentioned in \S \ref{secAutoCorDyn}. We show results for three representative parameters of the inner-most planet in each of these simulations ($P$, left column; $K$, middle column; $e$, right column). 
\label{figAcCompact}}
\end{center}
\end{figure*}

Figure \ref{figAcWide} shows the behavior of long-period near-resonant systems on AC$^{-1}(0)$. 
Once again, we see weak or no trends with $n_{chains}$ and the noisy behavior of $\sigma_\gamma$. 
The most significant trend is for AC$^{-1}(0)$ to increase when transitioning from a system near a second-order resonance (3:1) to a first-order resonance (either 2:1 or 3:2). 
It appears that RUN DMC mixed more rapidly for the 3:2 than the 2:1 system. 
We do not suspect this would be the case for all ranges of masses and orbital configurations, as we will show in Figure \ref{figAcCompact}. 
The takeaway from this figure is that systems near the 3:1 resonance has an AC$^{-1}(0)$ range comparable to the Keplerian two-planet system in Figure \ref{figAcAll}.
While RUN DMC is perfectly capable of computing posterior samples for strongly interacting systems, we find a larger autocorrelation of states from the Markov chains for systems near first-order resonances (2:1 or 3:2), implying that a significantly larger number of generations will be needed when modeling such systems.

In Figure \ref{figAcCompact}, we consider closely separated planet pairs, which undergo many cycles on the observing timescale. 
The simulated short-period 3:1 system shows a correlation with AC$^{-1}(0)$ being insensitive to $n_{chains}$ and $\sigma_\gamma$.
For the 3:1 system, the absolute values of AC$^{-1}(0)$ are comparable to the 3:2 long-period system.
As we ramp up the strength of the interactions, we see a steady increase in AC$^{-1}(0)$ for all the parameters considered.
In the worst case, a 3:2 short-period system requires over 1000 generations to get effectively independent samples.
Combined with Figure \ref{figAcWide}, the general conclusion from our results is decreasing the orbital separation of planet pairs (which typically boosts the strength of the planet-planet interactions) tends to increase the difficulty of sampling the posterior distribution.

\section{Discussion}
\label{secDiscuss}
We have performed several thousand RUN DMC simulations to test the algorithm's robustness and efficiency. 
We find that the DEMCMC algorithm performs very well for the analysis of realistic RV datasets, including systems with observable mutual planetary interactions.  
The algorithm can recover reliably from an initial ensemble of states that deviate significantly from the target posterior density.

Based on our results, we can provide general recommendations for the choice of $n_{chains}$, $n_{gen}$, and $\sigma_\gamma$ in future applications to real RV data. 
We find that both the minimum number of generations required for burn-in and the posterior sampling efficiency are most sensitive to the $n_{chains}$ parameter. 
The optimal value for $n_{chains}$ varies with $n_p$. 

We found no significant trend with $\sigma_\gamma$, but large values ($\ge0.1$) sometimes hindered parameter space exploration. 

Assuming a fixed $n_{eval}=n_{chains}\times n_{gen}$, we recommend using a large $n_{gen}$ over a large $n_{chains}$, subject to a firm lower limit of $n_{chains}>n_{dim}$ based on the geometry of drawn proposal vectors of the DEMCMC algorithm. 
For some data sets, we observed inefficient sampling when $n_{chains}$ is only slightly larger than $n_{dim}$. 
A few tens of chains work best for two-planet systems, but four-planet systems may require a couple of hundred chains. 
Thus, we recommend choosing $n_{chains}\sim 3 n_{dim}$, a value consistent with the recommendation from \cite{terBraak06} for a unimodal posterior, unless one uses a highly parallel architectures (e.g., GPU or cloud) where the wall clock time does not scale linearly with the number of model evaluations. 

In summary, we recommend:
\begin{itemize}
\item $n_{chains}\sim 3 n_{dim}$
\item $\sigma_\gamma < 0.1$
\item Large $n_{gen}$ is favored over than large $n_{chains}$
\end{itemize}

Considering the burn-in phase, we found that for initial conditions based on a shift perturbation of a fixed magnitude ($\beta$), the burn-in time increased with $n_p$. 
The one notable exception was that RUN DMC struggled with the $\mu$ Ara planetary system.  
Follow-up simulations demonstrate that primary challenge is not sampling such a high dimensional parameter space but rather the relatively sparse RV dataset providing looser estimates of our model parameters.
For this system, only $\sim$50\% of the states in the final generation were near the global mode after the first $\sim$1000 generations. 
In the simulations where some states were still far from the global posterior mode, we observed that they were typically scattered widely about parameter space, had significantly larger $\chi_{eff}^2$ than the samples near the global mode and were not clumped together in a single local minimum. 
Thus, they are easily recognized and can be replaced with samples from previous generations to provide an improved initial ensemble for a second RUN DMC simulation that can be used for inference, even in challenging cases such as $\mu$ Ara.
Overall, we recommend that the following for an efficient burn-in phase:
\begin{itemize}
\item For two and three-planet systems, a burn-in of at least $\sim$several hundred and roughly 1000 generations, respectively, are required. 
\item For four-planets with considerably more observations, we recommend a burn-in of at least 1000 generations, while avoiding $n_{chains}<64$.
\end{itemize}

%
%

Our results also provide insight into how long a RUN DMC simulation must run to eventually obtain enough independent samples to estimate model parameter uncertainties accurately. 
In general we find that the lag needed to achieve an AC$\approx$0 increases gradually with $n_{p}$, with the most dramatic increases occurring for systems where planet-planet interactions are significant during the timespan of observations.  

Clearly, multiple variables (number of observations, signal-to-noise of each planet, orbital architecture) affect the navigability of the respective posterior distribution for this (or other) sampling algorithms. 
While the vast majority of currently available dataset could be analyzed in a nearly automated fashion, analyzing some of the most complex and challenging RV data sets still require an expert's physical and statistical intuition.
The values we obtained for our various metrics provide the foundation future RV work using DEMCMC.  
Nevertheless, the dynamics of near MMR systems can be sufficiently complex, that we recommend a careful analysis of each systems on an individual basis.

Extrapolating to higher dimensionality, we can make predictions for how much computational effort is needed to analyze systems with 5+ planets. 
For example, we analyzed a simulated four-planet system based on 55 Cnc, as a precursor to a future, more detailed study of the system.
Despite the strong planet-planet interactions, the long observing baseline, and a wide range of RV signal-to-noise for different planets in the system, RUN DMC quickly produced an ensemble of states near the global posterior mode. 
Based on our results, we anticipate needing at least $n_{chains}\sim64$ and 1,000 generations for burn-in, followed by $\sim$3000 $\times$ $n_{goal}/n_{chains}$ generations of posterior samples.
Since at least two of the planets are likely strongly interacting, we expect that the required burn-in time may increase substantially. 
Indeed, this paper lays the foundation for a series of papers with detailed analyses and scientific interpretation of specific systems.
We intend to apply RUN DMC to most of the multiple planet systems with strong planetary interactions and publicly available radial velocity sets.
We are working towards applying the algorithm to even more challenging systems, such as 55 Cnc (B. Nelson et al. in prep.) and GJ 876 (B. Nelson et al. in prep.).

\acknowledgements 
We thank the anonymous referee for his or her very helpful comments that helped strengthen the paper.
This research was supported by NASA Origins of Solar Systems grant NNX09AB35G, NASA Applied Information Systems Research Program grant NNX09AM41G, NASA Kepler Participating Scientists Program grants NNX09AB28G and NNX12AF74G, and NASA Origins of Solar Systems Program grant NNX13A124G.
The authors acknowledge the University of Florida High-Performance Computing Center for providing computational resources and support that have contributed to the results reported within this paper.  

This material was also based upon work partially supported by the National Science Foundation under Grant DMS-1127914 to the Statistical and Applied Mathematical Sciences Institute.
Any opinions, findings, and conclusions or recommendations expressed in this material are those of the author(s) and do not necessarily reflect the views of the National Science Foundation.

\bibliography{references}

\clearpage

\end{document}